\documentclass[a4paper,twoside]{article}

\usepackage{epsfig}
\usepackage{subfigure}
\usepackage{calc}
\usepackage{multicol}
\usepackage{pslatex}
\usepackage{fancyhdr}
\usepackage[small]{caption}

\usepackage{epsfig}
\usepackage{alltt}

\def\beq{\begin{eqnarray}}
\def\eeq{\end{eqnarray}}
\def\2{\frac{1}{2}}
\def\sex{\begin{alltt}\small}
\def\esex{\end{alltt}}

\newtheorem{definition}{Definition}

\newcommand{\promise}[1]{\stackrel{\pi:#1}{\longrightarrow}}

\newcommand{\bundle}[1]{\stackrel{\Pi:#1}{\Longrightarrow}}

\title{Program Promises}

\author{{\textmd {\textbf }}\\
  \fontsize{11}{13}\selectfont
  Demissie Aredo, Mark Burgess and Simen Hagen\\
  \fontsize{9}{11}\selectfont
  \textit{Department of Computing, Oslo University College, Norway}\\
  \fontsize{9}{11}\selectfont
  \textit{Demissie.Aredo@iu.hio.no, Mark.Burgess@iu.hio.no, Simen.Hagen@iu.hio.no}\\[12pt]
}

\date{June 2007}

\begin{document}

\maketitle

\begin{abstract}
{\fontsize{9}{11}\selectfont 
The framework of promise theory offers an alternative way of
understanding programming models, especially in distributed
systems. We show that promise theory can express some familiar
constructs and resolve some problems in program interface design, using fewer and
simpler concepts than the Unified Modelling Language (UML).}
\end{abstract}

\section{\uppercase{Introduction}}

Current methodology in computer programming is based mainly on a popular
interpretation of Object Orientation (OO) and software architecture
modelling is usually done using the Unified Modelling Language
(UML) \cite{rjb2004}. Finding the right balance between top-down and
bottom-up in program design is no easy matter, but OO and UML
encourage programmers to begin by describing a taxonomy of classes (a
high level concept) for use as low level data types.  This approach
can be criticized in a number of ways. First, the resulting tree
structures are only a subset of the possible topologies that a program
{\em could} have, and the implications of the restriction are
unclear. Secondly, focusing on classes before a clear understanding of
how they will be used as instances is obtained can lead to mistaken
assumptions. In this paper we consider whether there is a natural
program structure from the viewpoint of promise theory, a theory of
declarations about what can happen between interacting components (as
opposed to what is assumed to happen in particular use-cases).  In
such a short paper, it is not possible to fully justify such an idea,
but we believe that the great simplicity of promise theory has virtue
that is worthy of further study. The resulting viewpoint leads to a
form of Service Oriented Architecture (SOA).

\section{A lightning introduction to promise theory}

Promise theory is a high level, graph-theoretical description of
`agents' which exhibit constrained behaviour. These agents are
completely independent entities and they document the behaviours they expect to
exhibit by making promises.
Programming objects, functions and data
will all be agents in our story. Agents in promise theory are truly
autonomous, i.e. they cannot be forced into performing any service of
behaviour by external agents. Instead they voluntarily cooperate with one
another \cite{burgessDSOM2005}. This idea allows us to easily model unreliability
and related issues by adding probabilities to promise graphs \cite{siri1}.
Consider two agents $a_1$ and $a_2$, where the first agent wishes to promise
that it will behave according to a behavioural specification $b$.
\begin{definition}[Promise]
An autonomous statement of, as yet, unverified behaviour, written:
\beq
a_1 \promise{b} a_2.
\eeq
where $b$ is the ``promise body''.
\end{definition}
Agents are completely impenetrable to outside influence, they have private
knowledge, and the promises that they make to one another cannot be
coerced. A promise with body $+b$ is understood to be a specification
to ``give'' behaviour from one agent to another (possibly in the
manner of a service), while a promise with body $-b$ is a
specification of what behaviour will be ``received'' or ``used'' by
one agent from another (see table \ref{summary}).  A promise {\em
valuation} $v_i\left(a_j\promise{b} a_k\right)$ is a subjective
interpretation by agent $a_i$ (in any appropriate currency) of the
promise in the parentheses. Usually an agent can only evaluate
promises in which it is involved.

\begin{table}[ht]
\begin{center}
\begin{tabular}{c|l}
\hline
\hline
Symbol & Interpretation\\
\hline
$a \promise{+b} a'$ & Promise with body $b$\\
$a' \promise{-b} a$ & Promise to accept $b$\\
$v_a(a \promise{b} a') $ & The value of promise to $a$\\
$v_{a'}(a \promise{b} a') $ & The value of promise to $a'$\\
\hline
\end{tabular}
\bigskip
\caption{Summary or promise notation\label{summary}}
\end{center}
\end{table}

Promises can be made about any subject that relates to the behaviour
of the promising agent, but no agent can make promises about another's
behaviour.  The subject of a promise is represented by the
promise body $b$, which consist of two essential parts: a promise {\em
type} ($\tau$) and a {\em constraint} ($\chi$) which indicates what
subset of behaviours are promised from within the domain of all
behaviours of that type.  Finally, conditional promise bodies
are written $b/c$, meaning that the body $b$ is promised if the
conditional $c$ is promised.

Promise theory is `simpler' than the Unified Modelling Language, in
that it has only one type of diagram (a labelled graph) and some
simple algebraic rules \cite{siri1}, however many details are moved
from diagrams into the definition of promise types and others are
simply suppressed). This simplicity offers clarity and the discipline
of voluntary cooperation reveals problems that are not apparent in
obligation models of UML, as we show below.

We define the
concepts of a bundle and a role of promise types.  We shall use this
concept below to define a {\em class} in the object oriented sense.

\begin{definition}[Promise bundle]
Let $S,R \subseteq A$ be arbitrary subsets of equal dimension $d$, from the
set of agents. A promise bundle is any collection of promises made (one-to-one)
from agents in $S$ {\em onto} agents in $R$. We denote this
\beq
S \bundle{B} R\; \equiv\; s_1 \promise{b_1} r_1, \ldots s_d, \promise{b_d} r_d.
\eeq
where $\Pi$ denotes a bundle of promises and 
$B$ is the collection of promise bodies $b_1,b_2,\ldots, b_d$.
\end{definition}
In promise theory, roles are associated both with {\em types} of
promises $\tau$ and their topology. In this respect, we can associate a
collection of agents with a role, given that they make (or use) the
same set of promise types.

\begin{definition}[Role]
Let ${\mathcal A}$ be a set of autonomous agents.  The family of sender/receiver
subsets $S,R\subseteq {\mathcal A}$ in a promise graph, which take part in
promises with body types $t(B)$, define $B$-roles.  Any subset $S$ or $R$
in the graph is said to play a role of (sender/receiver) with body types $t(B)$ in
the promise graph. Bundles with equal types and different constraints 
have equal roles.
\end{definition}
Promise roles are identified empirically (as `behaviour'
must be), not `designed' apriori. For instance,
the set of all agent nodes $W \subseteq {\mathcal A}$ that promises to
provide web service to any agent $R \in {\mathcal A}$: $W \promise{\rm
web} (R \in {\mathcal A})$ plays a role, which we can call `web
server'.  The set of nodes $C \subseteq {\mathcal A}$ that is promised
web service by any agent: $(S \in {\mathcal A})\promise{\rm web} C$
plays a role, which we can call `web client'.  We refer readers to
 \cite{siri1,siri2} for more information about promises and roles.

\section{\uppercase{Programming concepts}}\label{oop}

Modern programming methods rely heavily on compound types like
`record' or `struct'.  These are bundles of primitive data-types,
hence bundles of promises about values. A class can additionally
include methods as members.  To save space, we state without proof an
equivalence between methods and data values in promise theory. Since
one can always write a method that simply returns a data value, we
hope this is fairly intuitive.  Using this equivalence of
values and methods with promised services, we can now reinterpret
OO-classes as collections of promise bodies from similar promise
bundles.

\begin{figure}[ht]
\psfig{file=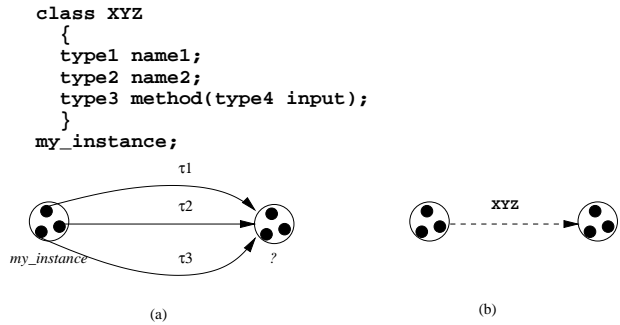,width=8cm}
\caption{\small A class instance is a promise bundle with a unique pattern.\label{class}}
\end{figure}

It is straightforward to map the structured class name tuples in
fig. \ref{class} to a flat promise type space. For instance, from
figure \ref{class}:
\beq
\langle XYZ, type1, name1 \rangle &\rightarrow& \tau_1,\nonumber\\
\langle XYZ, type2, name2 \rangle &\rightarrow& \tau_2,\nonumber\\
\langle XYZ, type3, method(input) \rangle &\rightarrow& \tau_3/\tau_4\nonumber\\ ~~etc.
\eeq
The hierarchical structure is not relevant to us; we only need to
uniquely identify the members.
  It is thus trivial to model {\em class instances} or objects
as bundles of promises that are isomorphic (see fig. \ref{class}).
One of the controversial aspects of Object Oriented modelling is the
idea that classes model real world situations and therefore lead to
simpler programs. Let us take a different view, motivated by the idea
of promises. Suppose we describe a program in terms of a set of
promises it must keep: let us ask, is there a natural set of promise
patterns that emerge in the promise graph of the program? Such basic
patterns might be a natural set of classes for the program, or at
least form a minimal {\em spanning set}. To explore this idea, we must
establish a mapping between promises and OO concepts.

We assume here, for simplicity, that we are operating in a
region of common data types and map promise types to data types.
\begin{definition}[Class-like promise body]
Any unique collection of promise bodies $c \equiv \{b_1,b_2,\ldots\}$,
belonging to a promise bundle.
\end{definition}
A class-like bundle is a collection of promise bodies, i.e. it is
missing the sender and the receiver. It is thus an abstract compound
type.

In other words, class-like instances form distinguishable patterns of
promises, that occur possibly several times in the graph and they must
agree on the basic type-alphabet for their pattern matching, on a
per-promise basis.  A class-like bundle could now naturally be
associated with an OO class, and the OO-class is thence the role
belonging to a promise bundle. However, we must remember that the
naming of classes in OO is a user policy decision, not a
requirement. Thus, we claim that while the roles lead us to a
`natural' set of classes, there is no obligation for any programmer,
agent or analyst to adopt these as the actual set of OO classes.  What
we have identified is a natural `spanning set' of class containers. This
emphasizes that OO is a policy, not a necessity.

\subsection{Inheritance and usage}

The concept of inheritance is central in Object Orientation, but it is
often ambiguously described. It is both used and explained using a number
of different
interpretations \cite{snyder86encapsulation,lieberman86using,cook89denotational}.
We can use promises to restate the meaning of inheritance more
clearly, though we cannot claim that our interpretations are those
intended by other authors.  We shall consider three such meanings:

\begin{itemize}

\item {\em Class 2 extends (adds to) class 1:}

When a child class extends another parent class, it adds values and methods
that were not present in the parent class.

\item {\em Class 2 substitutes (replaces) class 1}

When a child class substitutes a parent class, it replaces
it completely with a new implementation. In the case of an
abstract parent class it might, in fact, provide the first proper
implementation.

\item {\em Class 2 specializes (overrides) class 1:}

A subset of values and methods in the parent class can be
replaced with new implementations. In the limit of complete
overriding, this becomes the same as substitution.

\end{itemize}
Programming languages generally lump all three inheritance meanings
together in a general construct, so an inheritance policy might imply
several of these at the same time (e.g. one often extends a parent
class into a number of child classes which are then thought of as
specializations -- here we would call them mutually exclusive
extensions). We wish to be somewhat clear about these and therefore
keep them separate.  In addition to this there is {\em usage} or {\em
delegation} in which a class ``out-sources'' functionality to other
classes by aggregation of multiple specialized classes. This is
similar in spirit to extension, but does not attempt to
``philosophically'' integrate the functionality under the umbrella of
the same class.

The choice of inheritance and delegation is a policy choice, often
motivated by a heuristic philosophy.  The Liskov Substitution
Principle (Class 2 ``IS A'' Class 1) has often been used to explain
when inheritance should be used over delegation, for
instance \cite{liskov93family}. However, the ``is a'' relationship is
itself a heuristic one, in which it is unclear whether one is talking
about syntax or behaviour.  This leads to ambiguities in its
application.
Inheritance is thus used with several inequivalent meanings in OO, and
so we must disentangle the semantics of these distinct policies as
behavioural promises.

{\em Extension} is a relationship that decides whether a parent class is a subset
of the derived child class. Let us define these in terms of promises.
\begin{definition}[Extension]
Let $c$ and $c'$ be two class-like bundles,
We say that $c$ extends $c'$
or $c > c'$
iff $c' \subseteq c$. This gives us the notion of a container, and containment.
\end{definition}
Notice that this is simply a rule for aggregation. In terms of promises,
there is no difference between the extension of a class and the formation
of a bundle.

{\em Overriding} promises (methods and values) is typically a technique used
in sub-type polymorphism.  When overriding, there cannot be a conflict
between parent and client by definition, since the new behaviour and
the old behaviour are mutually exclusive. Overriding requires a
switch-like structure (indeed subtype-polymorphism is often intended
to replace switch-case constructions). Let $S$ and $R$ be agents,
promise bundles can be selected and hence overridden based on type
labels using a construction in which bundles are made
exclusively if a certain type is promised. Since types are mutually
exclusive, the promises are mutually exclusive:
\beq
S &\bundle{c_1/{\rm type=type1}} &R\nonumber\\
S &\bundle{c_2/{\rm type= type2}} &R\nonumber\\
R &\promise{\rm type} & S\nonumber\\
S &\promise{U({\rm type})} & R
\eeq

{\em Specialization} is a selective use of overriding to alter a subset of 
promises.
\begin{definition}[Specialization]
Let $c_p$ and $c_i$, $i=1\ldots N$ be class-like bundles.
We say that $c_i$ specializes $c_p$
iff a subset $c \subseteq c_p$ of class promises in the bundle is made
conditionally on predicate $p$ and all promises in $c_i$ are
dependent on predicates $p_i$, in such a way that $p$ and $p_i$ are
all mutually exclusive. It is normally assumed that the promise types
in $c_p$ and $c_i$ are identical.
\end{definition}
Thus we define a polymorphic class replacement using a simple
switch-case construction as one would expect.

For example, consider promise bundles that belong to the parent and
the child.  These are conditionally switched if the promise of a
subtype is made by the receiving agent.
\beq
S &\bundle{c_{\rm base\_parent}} &R\nonumber\\
S &\bundle{c_{\rm ext\_parent}/\neg{\rm subtype}} &R ~{\rm (optional)}\nonumber\\
S &\bundle{c_{\rm child}/{\rm subtype}} &R\nonumber\\
R &\promise{{\rm subtype}} &S\nonumber\\
S &\promise{U({\rm subtype})} &R
\eeq
Here the extension-parent class bundle is exchanged for the child
bundle if subtype is promised. The base-parent bundle is promised in
both cases.  If a parent class wishes to enforce behavioural
constraints on a child, then it must make such base-promises that are
not overridable, otherwise there can never be any broken promises
in spite of radically changing behaviour.

Specialization is a selective use of overriding to alter a subset of 
promises.
\begin{definition}[Substitution]
Let $c_p$ and $c_i$, $i=1\ldots N$ be class-like bundles.
We say that
 $c_i$ substitutes $c_p$
iff the body $c_p$ in the bundle is made
conditionally on some predicate $p$ and all promises in $c_i$ are
dependent on predicates $p_i$, in such a way that $p$ and $p_i$ are
all mutually exclusive. It is assumed that $c_p$ and $c_i$ are
isomorphic with respect to promise types and number, so that
every $c_i$ is a complete replacement for $c_p$.
\end{definition}
Through these definitions we have unwittingly shown how to implement
ontological mappings for OO within the Service Oriented Architecture (SOA)
through type agreement. This is not something easily represented in UML.

\section{\uppercase{The Liskov test}}

Our notion of substitution is stronger than the weak form often used
in the literature. The Liskov substitution test 
 \cite{liskov93family} was originally proposed as a syntactic and
semantic guideline on ``good inheritance'' and attempted to summarize
and make sense of the practice of inheritance by introducing a test of
class compliance.  The test has since often been represented by the
introduction of the heuristic ``is a'' relation, whose name seems to
imply an equality of certain entities, but this is not the
case. Unfortunately this term is rarely given a clear definition and
therefore leads to confusions and apparent paradoxes.  We shall make
simple sense of this relation with our own definition as follows:

\begin{definition}[``is a'']
A class-like promise bundle $S \bundle{c} R$ ``IS A'' class-like
promise bundle $S \bundle{c'} R$ iff both bundles can be promised
at the same time without breaking any promises.
\end{definition}

An example that is frequently used to illustrate an apparent paradox
with this is that of a rectangle and a square.  Clearly a square is a
special case of a rectangle as a concept, in every mathematical sense,
however a square has fewer behavioural freedoms than a rectangle (not
more) so it cannot extend the notion of rectangle in behavioural
terms. This indicates that extension is often at odds with the notion
of specialization. We can see this as follows.  Let $R$ be a rectangle
agent, $S$ be a square agent and $A$ be any other promisee agent.
\beq
R& \promise{{\rm width}=w }& A\nonumber\\
R& \promise{{\rm height}=h}& A\nonumber\\
R& \promise{{\rm angle}=90}& A\nonumber\\
R& \promise{{\rm sides}=4} &A
\eeq
We assume they have a common data-type schema, so we let square
extend rectangle:
\beq
S& \promise{{\rm width}=w }& A\nonumber\\
S& \promise{{\rm height}=h}& A\nonumber\\
S& \promise{{\rm angle}=90}& A\nonumber\\
S& \promise{{\rm sides}=4}& A\nonumber\\
S& \promise{w=h}& A
\eeq
The only difference now is that an additional promise has been made.
This is a behavioural constraint (not easily implementable in a
programming language, but straightforward in terms of promises).
Now, since the promises
\beq
S& \promise{{\rm width}=w }& A\nonumber\\
S& \promise{{\rm height}=h}& A\nonumber\\
S& \promise{{w=h}}& A
\eeq
are not independent, they may be reduced to:
\beq
S& \promise{{\rm width}=h}& A\nonumber\\
S& \promise{{\rm height}=h}& A.
\eeq
And if we now try to ask the question whether square ``is a'' rectangle,
it implies that all promises of $S$ and $R$ must hold simultaneously.
However, we quickly find a broken promise:
\beq
S/R &\promise{{\rm width}=w}& R\nonumber\\
S/R &\promise{{\rm height}=h}& R.
\eeq
Hence a square ``is not a'' rectangle in terms of the behaviours we
have attributed to it.  To avoid this problem one can simply introduce
a programming policy, ``Parent bundle promises should never be broken
by a child bundle''.

\section{\uppercase{A `natural' modelling example}}

Consider now a scenario in which the Unified Modelling Language (UML)
leads to a problematic solution, with little guidance to see what is
going wrong \cite{demissie}.  Consider a bank, comprised of a number of
employees and accounts for customers. Bank customers should have
access to use their accounts and bank employees are able to make
certain privileged transactions on accounts. If some
customers are in fact bank employees, they should not have privileged
access to perform transactions that include their own accounts (since
this could lead to embezzlement). Any user/Person, whether the owner
of an account or not should be able to make a cash payment to any
account. This example is somewhat simplistic, but therefore also
uncomplicated; however, it illustrates the basic conundrum.

\begin{figure}[ht]
\begin{center}
\psfig{file=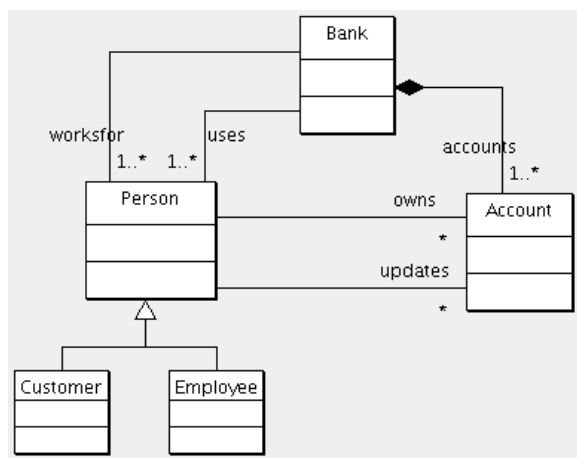,width=7.5cm}
\caption{\small An OO class model for the bank account/user interaction, in UML.\label{dem}}
\end{center}
\end{figure}

A not-unnatural UML model of this scenario is shown in
fig. \ref{dem}. We introduce the parent class ``Person'' from which
two child classes are derived by inheritance.  UML makes {\em
associations} between classes and in UML a subclass inherits
associations, as these form a subset of general attributes. Given only
this guidance, one could easily be led to the model shown in
fig. \ref{dem}
This structure has two problems that are easily avoidable by
reorganizing the model, but which are not ruled out by UML:
\begin{enumerate}
\item It assumes that customers and employees are mutually exclusive categories.
\item It allows employees to perform privileged updates on their own accounts.
\end{enumerate}
Although these are undesirable properties UML semantics do not offer
sufficient guidance to see what is going wrong.

Now consider a promise approach.  We introduce agents for users, bank
accounts and also a single (optional) agent representing the bank's
administration.  Always bearing in mind that agents in promise theory
cannot be {\em forced} to perform any function, and have limited knowledge,
we identify the necessary promises from Users to Account agents. We
further introduce a category attribute for users of accounts:
``customer'', ``employee'' and the default ``other''. The categories
``customer'' and ``other'' are disjoint.
We assume one account per agent and one user/Person per agent.  To
transport the necessary information to make promise-theoretic
decisions, a user/Person-agent must identify itself and its category. The
{\em possible} set of promises that are common to all agents are:
\beq
{\rm Person} &\promise{name=identity} & {\rm Account} ~~ (general)\nonumber\\
{\rm Person} &\promise{cash~payment}  & {\rm Account} ~~ (general)\nonumber\\
{\rm Person} &\promise{customer}      & {\rm Account} ~~ (subset)\nonumber\\
{\rm Person} &\promise{employee}      & {\rm Account} ~~ (subset)
\eeq
Some promises probably only make sense for certain categories of agent
and we might represent these as conditional promises:
\beq
{\rm Person} &\promise{use~account/customer}     & {\rm Account}\nonumber\\
{\rm Person} &\promise{priv~update/employee}     & {\rm Account}
\eeq
however, one could imagine that dishonest agents might try to make
these promises unconditionally. It is entirely up to the receiving
account to deal with such behaviour, according to the promise
precepts, so we shall disregard the conditionals here, as they make an
unwarranted assumption about agent behaviour.  Note that an agent
which does not promise to be a customer or an employee can still exist
and make promises, such as cash payments to any account.  Privileged
updates, on the other hand, should only be promised by employees.

An account agent makes a number of promises to users in return; some
of these are of a general nature, such as accepting information
promised to the agent:
\beq
{\rm Account} &\promise{U(name=identity)}       & {\rm Person}\nonumber\\
{\rm Account} &\promise{U(employee)}       & {\rm Person}\nonumber\\
{\rm Account} &\promise{U(customer)}       & {\rm Person}\nonumber\\
{\rm Account} &\promise{Keep~money~safe}       & {\rm Person}\nonumber\\
{\rm Account} &\promise{Account~functions}& {\rm Person}\nonumber\\
{\rm Account} &\promise{U(cash~payment)}  & {\rm Person}
\eeq
Some additional promises are conditional on the category of user, such as
the promise to accept instructions from different user categories.
This leads us naturally to an implementation in terms of access
control rather than class inheritance.
\beq
{\rm Account} &\promise{U(use~account)/C_1}& {\rm Person}\nonumber\\
{\rm Account} &\promise{U(priv~update)/C_2}& {\rm Person}
\eeq
Let us consider these conditions.
Any agent can make a cash payment to an account, regardless of
privilege level. For other transactions, we want to ensure
authorization however. The owner of an account should have all normal
usage privileges, thus if an agent with a name
matching.  The conditions $C_1,C_2$ must be based on
information that is available (promised) to the account agents:
\beq
C_1 : {\rm name} = owner ~~{\rm AND~~ employee} \not= true\nonumber\\
C_2 : {\rm name} \not= owner ~~ {\rm AND~~ employee} = true
\eeq
These conditional promises are mutually exclusive and hence they should
be represented by exclusive promises, which {\em extend} the
unconditional promises to accept cash payments.

We allow user/Person agents to be both customers and employees
non-exclusively, since the conditional tests can be made based
entirely on the information that a user is operating in employee mode
(e.g. making that promise by entering an authorization code or
passwd). We see from the services provided by an account that users
are distinguished entirely by the functions they carry out, not by
named skills without a basis in actual behaviour.

Account-agents promise to accept normal usage transactions as long as
an agent is not an employee (any agent customer or not might need to
pay money into a customer's account). An agent who is {\em sometimes
an employee} can also use its account as long as it is not currently
making the employee promise. As soon as employee privilege is invoked,
account agents invoke different promised behaviour which restricts the
actions they can perform.

What about the UML ownership association?  The account attribute
``owner'' is the private knowledge of the account-agent
concerned. There is no need for this information to be issued as an
explicit relation that is promised to a user, since that behaviour is
implicit in the Use-promises. Thus the ``owns'' association is not a
necessary interface relation, only an attribute of the account.

Readers might be uncomfortable that a bank account somehow knows
ownership information by itself -- this is not how many
would view a traditional bank.  If we wish to model a more
traditional, centralized bank, we could introduce a central bank agent
which has knowledge of all accounts and delegates the data to its
subordinate account-agents:
\beq
{\rm Bank} \promise{Name_i} {\rm Account}_i, ~\forall i
\eeq
Similarly, the account-agents could place themselves under the umbrella of the
bank by promising to use these names from the central bank agent.
\beq
{\rm Account}_i &\promise{U(Name_i)} &  {\rm Bank}\nonumber\\
{\rm Account}_i &\promise{K_{name}=Name_i} & {\rm Bank}
\eeq
There is nothing in this model that requires this however.
If one were to implement a bank as a Service Oriented device, then
bank accounts could be created anywhere without a common umbrella. 
They do not have to belong to a single bank. Our model is rather too
primitive to make a detailed case for this here however.

\subsection{The roles and classes}

\begin{figure}[ht]
\begin{center}
\psfig{file=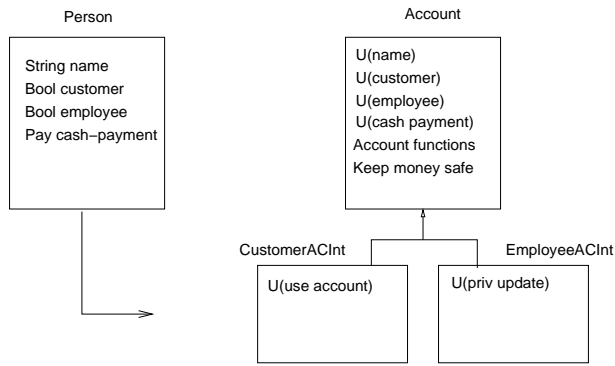,width=8cm}
\caption{\small A promise role derived class structure for the bank account problem. Note that the customer/employee roles are not represented as separate
classes, but as attributes to the common user (Person)
class.\label{sol}}
\end{center}
\end{figure}

We conjectured that the natural spanning set of classes in a system
can be discovered by looking for {\em roles} in the promise graph,
i.e. common and repeated patterns within the graph. When role-promises
are disjoint or mutually exclusive they fall naturally into separate
sub-classes, such as might be implemented through sub-type
polymorphism in OO. The solution below is shown in fig. \ref{sol}.

\begin{figure}[ht]
\begin{center}
\psfig{file=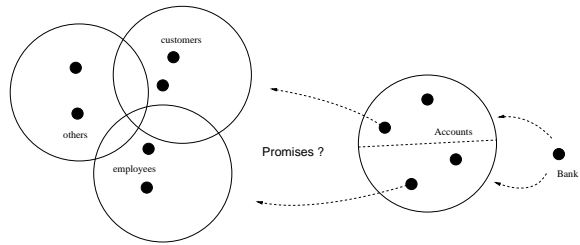,width=7.5cm}
\caption{\small Taking a promise viewpoint, the roles in the bank account interaction
example are shown. Note that different user roles overlap. An employee
might also be a customer in the bank, similarly user can pay in
transactions even when a user is not a customer of the bank.\label{roll}}
\end{center}
\end{figure}

We recognize three kinds of account promise roles, two of which are
mutually exclusive (and hence form natural subtypes) and one which is
common to the others (and hence requires no subtype). This forms a
natural tree decomposition as shown in fig. \ref{roll}.

What we see is that the promise theory solution, which makes behaviour
the premier consideration, is to split the account class into two
rather than the person/user class. The split is based on what promises
are available to the different kinds of users. The change to an
account subtype is a change in security level (like a pure
Clark-Wilson model \cite{clark1} interface) which leads to a natural
access control.
We end up with two security levels, implemented through mutually
exclusive interfaces (modes) in the account class, rather than two
kinds of users, since this arrangement embodies the behavioural
constraints.

\section{\uppercase{Discussion and conclusion}}

We do not know of any work truly related to the idea of promises,
though some superficially similar ideas exist in multi-agent theory
\cite{agents}. There is clearly a large literature on program
modelling but not from the perspective of voluntary cooperation.

Using our notion of promises, we are able to model some
salient features of program modelling and offer guidance to the
avoidance of common problems. There is an enormous potential for
describing interfaces and constraints on program activities, without
having to deal with actual computational processes.  Another
programming paradigm introduced recently is that of {\em aspects}.
Aspects allow one to weave in-line code fragments into an existing
program using a secondary compiler. The break with the hierarchical
container idiom makes this awkward to model in a class regime, but
this is easily represented as promises or services performed by a
weaving agent.

In OO design one models data-types first and creates instances of
these pre-decided types on demand as program logic develops. This
seems natural, since one does not necessarily know what precise
objects will be required in advance; however, it can lead to blind
alleys and refactoring, as one has no empirical basis for a project at
the outset.  A spanning set model, based on required objects and their
promises, which leads to `natural' organizational structures, can help
to avoid some of these blind alleys. We have applied this approach to
code examples in cfengine 3 \cite{cfwww} with some success.


Promises also allow analyses, both logical and economic.  The economic
trade-like model inherent in the SOA is naturally described in promise
theory.  Using the idea of common knowledge as a graph theoretic
construction (rather than a modal logic inference) one has an easy way
of tracking scope with graph theoretical tools \cite{knowledgemace2007}. This could be an
approach to examining {\em reflections}, or `self-aware' programs.
There are other ways entirely in which promises might assist in
modelling software engineering.  The act of basing software on a
specification is a promise to customers, saying that a program will
comply with its specification. On the other hand, much software is
developed without any formal specification. Does this make it less
able to function reliably?

What role do we foresee for promises in program modelling in the
future?  There are several possibilties. In cfengine 3, promises have
been made into a computer language for high level policy
declaration. It would be straightforward to add more user-friendly
graphical tools to rival the simplicity of UML, allowing policy to
encompass program structure more readily.  Promises clearly make SOA
thinking easier, so they are a natural tool for programmers there
(independently of web services). More importantly, we believe that
promises offer a flexible way of thinking that is at least as powerful
as hierarchical classes but which is less ``brittle'' and therefore
more robust to errors regardless of their source.

UML claims to model behaviour, but in fact it only describes assumed
internal transitions. Behaviour is something that must be observed
once a program is operating in its environment, and since the
environment is not modelled, behaviour is not modelled. This
distinction is clearer in promise theory and so there is some hope
that phenomena such as emergent behaviour can be understood in this
framework \cite{sirimace2007}.  We are developing these issues in other
work.

\section*{\uppercase{Acknowledgements}}

\noindent We are grateful to Jan Bergstra,
Alva Couch and Tuva Hassel Stang for stimulating discussions.
This work is supported by the EC IST-EMANICS Network of Excellence
(\#26854)


\bibliographystyle{unsrt}

\begin{thebibliography}{}

\bibitem[Aredo and Burgess, 2007]{knowledgemace2007}
Aredo, D. and Burgess, M. (2007).
\newblock On the consistency of distributed knowledge.
\newblock In {\em Proceedings of MACE 2007}, volume~6 of {\em Multicon Lecture
  Notes}. Multicon Verlag.

\bibitem[Aredo, 2004]{demissie}
Aredo, D.~B. (2004).
\newblock {\em Formal Development of Open Distributed Systems: Integration of
  UML and PVS}.
\newblock PhD thesis, Department of Informatics, University of Oslo, Norway.

\bibitem[Burgess, 1993]{cfwww}
Burgess, M. (1993).
\newblock Cfengine www site.
\newblock {\em http://www.cfengine.org}.

\bibitem[Burgess, 2005]{burgessDSOM2005}
Burgess, M. (2005).
\newblock An approach to understanding policy based on autonomy and voluntary
  cooperation.
\newblock In {\em IFIP/IEEE 16th international workshop on distributed systems
  operations and management (DSOM), in LNCS 3775}, pages 97--108.

\bibitem[Burgess and Fagernes, a]{siri1}
Burgess, M. and Fagernes, S.
\newblock Pervasive computing management: A model of network policy with local
  autonomy.
\newblock {\em IEEE Transactions on Software Engineering}, page (submitted).

\bibitem[Burgess and Fagernes, b]{siri2}
Burgess, M. and Fagernes, S.
\newblock Voluntary economic cooperation in policy based management.
\newblock {\em IEEE Transactions on Network and Service Management}, page
  (submitted).

\bibitem[Burgess and Fagernes, 2007]{sirimace2007}
Burgess, M. and Fagernes, S. (2007).
\newblock Laws of systemic organization and collective behaviour in ensembles.
\newblock In {\em Proceedings of MACE 2007}, volume~6 of {\em Multicon Lecture
  Notes}. Multicon Verlag.

\bibitem[Clark and Wilson, 1987]{clark1}
Clark, D. and Wilson, D. (1987).
\newblock A comparison of commercial and military computer security policies.
\newblock {\em Proceedings of the 1987 IEEE Symposium on Security and Privacy},
  page 184.

\bibitem[Cook, 1989]{cook89denotational}
Cook, W.~R. (1989).
\newblock A denotational semantics of inheritance.
\newblock Technical Report CS-89-33.

\bibitem[Lieberman, 1986]{lieberman86using}
Lieberman, H. (1986).
\newblock Using prototypical objects to implement shared behavior in
  object-oriented systems.
\newblock In Meyrowitz, N., editor, {\em Proceedings of the Conference on
  Object-Oriented Programming Systems, Languages, and Applications ({OOPSLA})},
  volume~21, pages 214--223, New York, NY. ACM Press.

\bibitem[Liskov and Wing, 1993]{liskov93family}
Liskov, B. and Wing, J. (1993).
\newblock {Family} {Values}: {A} {Behavioral} {Notion} {Of} {Subtyping}.
\newblock Technical Report MIT/LCS/TR-562b.

\bibitem[Rumbaugh et~al., 2004]{rjb2004}
Rumbaugh, J., Jacobson, I., and G.Booch (2004).
\newblock {\em Unified Modeling Language Reference Manual, The (2nd Edition)}.
\newblock Addison-Wesley Object Technology Series. Pearson Higher Education.

\bibitem[Snyder, 1986]{snyder86encapsulation}
Snyder, A. (1986).
\newblock Encapsulation and inheritance in object-oriented programming
  languages.
\newblock In Meyrowitz, N., editor, {\em Proceedings of the Conference on
  Object-Oriented Programming Systems, Languages, and Applications ({OOPSLA})},
  volume~21, pages 38--45, New York, NY. ACM Press.

\bibitem[Wooldridge, 2002]{agents}
Wooldridge, M. (2002).
\newblock {\em An Introduction to MultiAgent Systems}.
\newblock Wiley, Chichester.

\end{thebibliography}

\end{document}